\documentclass[amsmath,amssymb,aps,preprint,eqsecnum,prd]{revtex4}
\usepackage{graphicx}
\newcommand{\Ibar}{\emph{I}\kern-.24em\rule[0.72ex]{0.25em}{0.05ex}}
\begin{document}

\begin{flushright}
IGPG-04/11-1\\
gr-qc/0411030
\end{flushright}

\title[Freely precessing neutron stars]{The gravitational wave
  spectrum of non--axisymmetric, freely precessing neutron stars}

\author{Chris Van Den Broeck}
\email{vdbroeck@gravity.psu.edu}
\affiliation{Institute for Gravitational Physics and Geometry, Department
of Physics, The Pennsylvania State University, 104 Davey 
Laboratory, University Park, PA 16802}

\begin{abstract}
  Evidence for free precession has been observed in the radio
  signature of several pulsars. Freely precessing pulsars radiate 
  gravitationally at frequencies near the
  rotation rate and twice the rotation rate, which for rotation frequencies
  greater than $\sim 10$ Hz is in the LIGO band. In older work, the 
  gravitational wave spectrum of a precessing neutron
  star has been evaluated to first order in a small precession angle. 
  Here we calculate the contributions to second order in the wobble angle,
  and we find that a new spectral line emerges. We show that 
  for reasonable wobble angles, the second-order line may well be 
  observable with the proposed advanced LIGO detector for precessing neutron 
  stars as far away as the galactic center. Observation of 
  the full second-order spectrum permits a direct measurement
  of the star's wobble angle, oblateness, and deviation from axisymmetry,
  with the potential to significantly increase our understanding of
  neutron star structure.
  \vskip 0.5cm
  \noindent
  Pacs numbers: 04.30.Db, 97.60.Jd

\end{abstract}

\maketitle

\section{Introduction}

For some decades now, freely precessing neutron stars have been considered
as possible sources of detectable gravitational radiation. Early treatments
include the ones by Zimmermann et al.
\cite{zimmermann78a,zimmermann79a,zimmermann80a}
and by Alpar and Pines \cite{alpar85a}.
More recently interest in the subject was revived by compelling
observational evidence for free precession in the
radio signature of PSR~B1828-11 \cite{stairs00a} as well as 
PSR~B1642-03 \cite{shabanova01a}. Evidence for free 
precession had also been put forward for PSR~B0531+21 
(the Crab pulsar) 
\cite{jones88a,cadez97a} and PSR~B0833-45 (the Vela pulsar)
\cite{deshpande96a}. These examples suggest the existence of more such
objects, some of which might be viable sources for current or future
ground-based gravitational wave detectors such as LIGO.

Prompted by the evidence for free precession in PSR~B1828-11, 
Cutler and Jones \cite{cutler01a} revisited earlier work which
seemed to suggest that 
when a neutron star is treated as an elastic body, gravitational wave
back-reaction will quickly damp the precessing motion. Careful
consideration led them to conclude instead that in realistic 
precessing neutron stars, gravitational 
wave damping will play but a minor role in the star's evolution,
although there will be other sources of damping.
Jones and Andersson \cite{jones01a} then constructed a detailed 
model of freely precessing neutron stars and compared it with
the observational data. They concluded that the 
precessional interpretation of these observations required superfluid
vortex pinning of the star's crust to its core to be inhibited, thereby
challenging some of the then-prevalent ideas concerning a neutron star's 
liquid interior. Nevertheless, an analysis by Link and Epstein \cite{link01a} 
supported the precessional hypothesis for PSR~B1828-11. 

In \cite{jones02a}, 
Jones and Andersson evaluated the viability of precessing neutron stars as
sources for gravitational wave detection in the near future. 
To estimate the strength of the signal, they used the waveforms
computed by Zimmermann et al. \cite{zimmermann79a,zimmermann80a}. 
Their conclusions appear to imply that gravitational waves from 
precessing neutron stars are
unlikely to be detectable even with an advanced LIGO detector. However,
their paper was written before the more recent analysis of the expected
capabilities of LIGO II by Fritschel \cite{fritschel01a,fritschel03a},
which is the one we will use.

Here we re-examine the gravitational waves generated by rapidly 
rotating, precessing neutron stars, evaluating whether the radiation
from such sources might indeed be detectable, and what we may learn from their 
detection.

The motion of a freely precessing neutron star has
two characteristic frequencies associated with it: 
the precession frequency and the 
rotation frequency. The corresponding gravitational wave spectrum
consists of a series of discrete spectral lines, with a principal
line at or near a harmonic of the rotation frequency, surrounded 
by companion lines separated from the main line by multiples of
the precessional frequency.

In \cite{zimmermann80a}, Zimmermann  
evaluated to first order in small wobble angle the quadrupole 
gravitational radiation emitted by freely precessing, non-axisymmetric
neutron stars, approximated as rigid bodies. At first order the spectrum 
includes two spectral lines: one at approximately the pulsar frequency and 
another at its first harmonic. 
Unfortunately, knowledge of the first-order spectrum
yields little or no information about the physical characteristics
of the neutron star emitting the radiation. From an observational
point of view, this makes it imperative to study the spectrum in
more detail.

Our goal here is threefold. First, we want to 
show that it is reasonable to
take the precession angle as the primary expansion parameter among the
various `small' parameters that characterize the problem. Secondly,
we want to determine the conditions under which the second-order contributions
to the gravitational wave spectrum of a precessing neutron star are
observable.  Finally, we would like to explore what observation of these
second-order corrections can reveal about the physical properties of
the source.

Throughout this paper, we will use units where $G=c=1$ unless stated
otherwise. Wherever estimates are needed for the mass and radius of a
neutron star, we will use $M = 1.4 M_\odot$ and $R = 10^6$ cm. 
The neutron stars of interest to us are fast-spinning ones;
in estimates involving rotation frequencies we use $500$ Hz as
a reference.
We shall treat neutron stars as rigid bodies and limit ourselves
to the quadrupole approximation to the gravitational radiation. When
describing rigid body motion we will follow 
\cite{landau76a,zimmermann80a,abramowitz72a},
i.e., we use the physical conventions of \cite{landau76a} and the
mathematical notation of \cite{zimmermann80a,abramowitz72a}. In section
\ref{sec:review} we review the closed expressions for the
gravitational waveforms found in \cite{zimmermann80a} and in
\ref{sec:zimmerman} the first-order approximation of
\cite{zimmermann80a}. In section \ref{sec:2ndOrder}, the closed
expressions are used to construct a second-order expansion, and a new
contribution to the gravitational 
radiation spectrum of precessing neutron stars is
found. Section \ref{sec:detectability} first discusses the
detectability of all three spectral lines. We then explain how
physical information can be extracted from the second-order spectrum.
Finally, section \ref{sec:conclusions} provides a summary of the
results and conclusions.

\section{Gravitational radiation and precession}\label{sec:review} 

The motion of an isolated, freely precessing rigid body is most
conveniently described in terms of the Euler angles $\theta$, $\phi$
and $\psi$ that orient the body with respect to an inertial coordinate
system.  Following the conventions of \cite{landau76a}, denote the
coordinates of the inertial coordinate system centered on the body by $X$,
$Y$, and $Z$, with the coordinate vector $\widehat{e}_Z$ in the direction
of the body's angular momentum.  Similarly, denote the coordinates of
the body-fixed frame as $x$, $y$, and $z$, with coordinate vectors
$\widehat{e}_x$, $\widehat{e}_y$ and $\widehat{e}_z$ parallel to the
eigenvectors of the body's moment of inertia tensor, and with the
corresponding principal moments of inertia satisfying $I_z>I_y\geq
I_x$. 
The Euler angle $\theta$ is then the angle between
$\widehat{e}_Z$ and $\widehat{e}_z$, 
$\phi$ is the angle between $\widehat{e}_X$ and the line of nodes
that is the intersection of the $\widehat{e}_X \times \widehat{e}_Y$
plane and the $\widehat{e}_x \times \widehat{e}_y$ plane, while
$\psi$ is the angle in the $\widehat{e}_x\times\widehat{e}_y$ plane
between the line of nodes and $\widehat{e}_x$ (cf.\ \cite[figure
47]{landau76a}).  

Denote the time-dependent rotation matrix that transforms from the
body-frame to the inertial frame by $R_{j\mu}$, where Latin indices
refer to the inertial frame and Greek
indices refer to the body-fixed frame.
The quadrupole gravitational radiation from the precessing body is
proportional to the second time derivative of the moment of inertia
tensor in the inertial frame, which depends on the principal moments
of inertia and the time-dependent rotation matrix.

The standard quadrupole formula for the transverse-traceless (TT) gauge
metric perturbation is
\begin{equation}
h^{TT}_{jk} = \frac{2}{r}\frac{d^2\Ibar_{jk}}{dt^2}, \label{eq:htt}
\end{equation}
where
\begin{equation}
\Ibar_{jk} = \int d^3X\,\left(X_j X_k -
\frac{1}{3}{\bf X}^2\delta_{jk}\right)\,\rho({\bf X}),
\end{equation}
with $\rho({\bf X})$ the density of the source.
Eq.~(\ref{eq:htt}) can be written in terms of the rotation matrix
$R_{j\mu}$, the components 
$\Omega_\mu$ of the angular velocity in the body-fixed frame, and the
eigenvalues of the rigid body's moment of inertia tensor $I_1$,
$I_2$, and $I_3$ \cite{zimmermann80a}:
\begin{equation}  
h^{TT}_{jk}  = \frac{2}{r}R_{j\mu}A_{\mu\nu}R^T_{\nu{}k} \label{rewritten},
\end{equation}
where summation over $\mu$ and $\nu$ is understood.
$A_{\mu\nu}$ is given by
\begin{subequations}\label{A}
\begin{eqnarray}
A_{11} &=& 2 (\Delta_2 \Omega_2^2 - \Delta_3 \Omega_3^2), \\
A_{12} &=& \Gamma_3 \, \Omega_1 \Omega_2, 
\end{eqnarray}
\end{subequations}
with the other components being defined by symmetry and 
cyclic permutation of the
indices. $\Delta_\mu$ and $\Gamma_\mu$ are shorthand for
\begin{subequations}\label{deltagamma}
\begin{eqnarray}
\Delta_1 &=& I_2 - I_3,\\
\Gamma_1 &=& \Delta_2 - \Delta_3 + {\frac{\Delta_1^2}{I_1}},
\end{eqnarray}
\end{subequations}
where the other components are again given by cyclic permutation of the 
indices. 

To facilitate the description of the radiation, introduce two orthogonal unit
vectors $\widehat{v}$ and $\widehat{w}$ with
$\widehat{v}\times\widehat{w}$ directed toward a distant observer of
the emitted gravitational waves.  Without loss of generality we may
assume that the
observer lies in the $\widehat{e}_Y\times\widehat{e}_Z$ plane. Let $i$
denote the angle between $\widehat{e}_Z$ and the line of sight to the
observer, and define
\begin{subequations}
\begin{eqnarray}
\widehat{v} &=& \widehat{e}_Y\cos(i) - \widehat{e}_Z\sin(i),\\
\widehat{w} &=& -\widehat{e}_X. 
\end{eqnarray}
\end{subequations}
In this frame the TT gauge metric perturbation
corresponding to the observed radiation can be described in terms of
two scalar functions, $h_{+}(t)$ and $h_{\times}(t)$, representing
the radiation in the two gravitational wave polarizations:
\begin{subequations}\label{pluscross}
\begin{equation}
h^{\text{TT}}_{ij}(t) = h_{+}(t)\,(\widehat{e}_{+})_{ij}
+ h_{\times}(t)\,(\widehat{e}_{\times})_{ij},
\end{equation}
with
\begin{eqnarray}
\widehat{e}_{+} &=& 
\widehat{v}\otimes\widehat{v} -
\widehat{w}\otimes\widehat{w}, \\
\widehat{e}_{\times} &=& \widehat{v}\otimes\widehat{w} +
\widehat{w}\otimes\widehat{v}, \\
h_{+} &=& \frac{1}{2}h^{TT}_{jk}e^{jk}_{+},\\
h_{\times} &=& \frac{1}{2}h^{TT}_{jk}e^{jk}_{\times}.
\end{eqnarray}
\end{subequations}
Using equation (\ref{rewritten}) for the TT gauge metric perturbation we
have 
\begin{subequations}\label{eq:closed}
\begin{eqnarray}
h_{+} &=&-\frac{1}{r} 
\left[
\left(R_{y\mu}\cos(i) - R_{z\mu}\sin(i)\right)
\left(R_{y\nu}\cos(i) - R_{z\nu}\sin(i)\right) 
- R_{x\mu} R_{x\nu}
\right]A_{\mu\nu},\\
h_{\times} &=& 
\frac{2}{r}
\left(R_{y\mu}\cos(i) - R_{z\mu}\sin(i)\right)
R_{x\nu}A_{\mu\nu}. 
\end{eqnarray}
\end{subequations}

In these expressions time evolution arises from the time
dependence of the angular velocity in the body frame and the rotation matrix. 

When 
$J^2 \geq 2 E I_2$, with $J$ the magnitude of the angular momentum 
and $E$ the energy in the rigid body motion\footnote{When $J^2<2EI_2$ then the
  same equations, but with a consistent interchange of indices 1 and
  3, describe the motion.}, the components of the angular velocity in
the body frame are given by \cite{zimmermann80a}:
\begin{subequations}
\begin{eqnarray}
\Omega_1 &=& a\,\text{cn}(\tau,m),\\
\Omega_2 &=& a\left[
\frac{I_1\left(I_3-I_1\right)}{I_2\left(I_3-I_2\right)}
\right]^{1/2}\text{sn}(\tau,m),\\
\Omega_3 &=& b\,\text{dn}(\tau,m),
\end{eqnarray}
where $\tau$ is a rescaled time variable, 
\begin{equation}
\tau = bt\left[\frac{(I_3-I_2)(I_3-I_1)}{I_1I_2}\right]^{1/2},
\end{equation}
$\text{cn}$, $\text{sn}$, and $\text{dn}$ are
elliptic functions \cite{abramowitz72a}, and $m$ is given by
\begin{equation}
m = \frac{(I_2-I_1)I_1a^2}{(I_3-I_2)I_3b^2}.
\end{equation}
\end{subequations}
The constants $a$ and $b$ are such that at $t=0$, $\Omega_1 = a$ and 
$\Omega_3 = b$. When the precession angle is small, it will approximately 
equal $a/b$.

The rotation matrix $R_{j\mu}$ is given in terms of the Euler angles
$\theta$, $\phi$, $\psi$. The time dependence of these can in turn
be expressed in closed form in terms of elliptic and theta functions: 
\begin{subequations}
\begin{eqnarray}
\cos\theta &=& \frac{I_3b}{J}\text{dn}\tau,\\
\tan\psi &=& \left[\frac{I_1(I_3-I_2)}{I_2(I_3-I_1)}\right]^{1/2}
\frac{\text{cn}(\tau,m)}{\text{sn}(\tau,m)},\\
\phi &=& \phi_1 + \phi_2,
\end{eqnarray}
with 
\begin{eqnarray}
\exp\left[2i\phi_1(t)\right] &=&
\frac{
\vartheta_4\left(\frac{2\pi t}{T}-i\pi\alpha\right)
}{
\vartheta_4\left(\frac{2\pi t}{T}+i\pi\alpha\right)
},\\
\phi_2 &=& \frac{2\pi t}{T'},\\
\frac{2\pi}{T'} &=& \frac{J}{I_1} - 
\frac{2i}{T}\frac{\vartheta_4'(i\pi\alpha)}{\vartheta_4(i\pi\alpha)}.
\end{eqnarray}
The constants $\alpha$ and $T$ that appear above are given by 
\begin{eqnarray}
\text{sn}\left[2i\alpha K(m)\right] &=& \frac{iI_3b}{I_1a},\\
T &=& \frac{4K(m)}{b}
\left[\frac{I_1I_2}{(I_3-I_2)(I_3-I_1)}\right]^{1/2},
\end{eqnarray}
where
$\vartheta_4$ is the $4^{\text{th}}$ Jacobi theta function with nome
\begin{equation}
q = \exp\left(-\pi\frac{K(1-m)}{K(m)}\right), 
\end{equation}
\end{subequations}
and $K(m)$ is the complete elliptic integral of the first kind
\cite{abramowitz72a}. 

Within the weak-field, quadrupole approximation these closed
expressions fully describe the gravitational wave signal from a freely
precessing, rigid body. However, after all is said and done, the
problem involves just two periodicities determined by $T$ and $T'$. 
Accordingly, the radiation should be resolvable into
a discrete spectrum. Knowledge of the amplitude and frequencies of
the spectral lines greatly simplifies the analysis of
gravitational wave data. The discrete nature of the
spectrum becomes apparent when $h_{+}$ and $h_{\times}$, as given in
equations (\ref{eq:closed}), are expanded in terms of `small' parameters: 
the `wobble angle parameter' $a/b$, the rigid body's
oblateness, and its deviation from axisymmetry. Finding the second-order
contributions to the spectrum in these parameters, evaluating the
observability of the associated spectral lines, and exploring what 
can be learned from such observations will be our main considerations.

\section{First order contributions to the gravitational 
wave spectrum}\label{sec:zimmerman} 

In his first-order expansion of $h_+$ and
$h_{\times}$, Zimmermann identified two expansion
parameters, $m$ and $\delta$:
\begin{subequations}
\begin{eqnarray}
m &=&
\frac{I_2-I_1}{I_3-I_2}\frac{I_1}{I_3}\left(\frac{a}{b}\right)^2
\label{m},\\
\delta &=& 1 -
\left({\frac{I_1}{I_2}}{\frac{I_3-I_2}{I_3-I_1}}\right)^{1/2}. 
\label{delta}
\end{eqnarray}
\end{subequations}
To find the `first-order' contributions to the radiation spectrum,
Zimmermann 
\begin{itemize}
\item expanded the rotation matrix $R_{i\mu}$ to first order in $m$
  and $\delta$;
\item substituted the expanded rotation matrix into the relation
  between the inertia tensor in the inertial frame and that in the
  body frame;
\item substituted the inertia tensor in the inertial frame into the standard
  quadrupole formulae for the strains (\ref{eq:htt});
\item truncated the resulting $h_{+}$ and $h_{\times}$ to first order
  in small $a/b$, small oblateness $(I_3-I_1)/I_3$, and small
  non-axisymmetry $(I_2-I_1)/(I_3-I_2)$.
\end{itemize}
This leads to
\begin{subequations}\label{waveform} 
\begin{eqnarray}
h^Z_+ &=& 
-\frac{2I_3\Omega_{\text{rot}}^2}{r}
\frac{\left(I_2-I_1\right)}{I_3}\left(1+\cos^2(i)\right)
\cos(2\Omega_{\text{rot}}t)
\nonumber\\  
&& +
\frac{I_3\left(\Omega_{\text{rot}}+\Omega_{\text{prec}}\right)^2}{r} 
\frac{2I_3-\left(I_1+I_2\right)}{2I_3}
\left({\frac{I_1 a}{I_3 b}}\right) 
\sin(2i)\,\cos[(\Omega_{\text{rot}} + \Omega_{\text{prec}})t],\\
h^Z_\times &=& 
-\frac{4I_3\Omega_{\text{rot}}^2 }{r} 
\frac{(I_2-I_1)}{I_3}
\cos(i) \,\sin(2\Omega_{\text{rot}} t) \nonumber\\
 && +
\frac{2I_3\left(\Omega_{\text{rot}}+\Omega_{\text{prec}}\right)^2 }{r}
\frac{2I_3-(I_1+I_2)}{2I_3}
\left(\frac{I_1 a}{I_3 b}\right) 
\sin(i) \,\sin{}[(\Omega_{\text{rot}} +
\Omega_{\text{prec}})t].
\end{eqnarray}
\end{subequations}
The fundamental angular
frequencies in the problem are 
\begin{subequations}
\begin{eqnarray}
\Omega_{\text{prec}} &=& \frac{2\pi}{T}\nonumber\\  
                     &=& {\frac{\pi b}{2
                  K(m)}}\left({\frac{(I_3-I_2)(I_3-I_1)}{I_1 I_2}} 
                    \right)^{1/2}, \label{precfreq}\\
\Omega_{\text{rot}} &=& \frac{2\pi}{T'}-\frac{2\pi}{T} \nonumber\\ 
                     &=& {\frac{J}{I_1}} - \left(1+{\frac{i}{\pi}}   
         {\frac{\vartheta'_4(i\pi\alpha)}{\vartheta_4(i\pi\alpha)}}\right)\,
           \Omega_{\text{prec}}.  
\label{rotfreq}
\end{eqnarray}
\end{subequations}

The expressions (\ref{waveform}) represent Fourier expansions of the
waveforms, with gravitational spectral lines at the angular frequencies
\begin{equation}
\begin{array}{rclcl}
\Omega^I &=& 2\Omega_{\text{rot}} &\qquad &\text{(line I)}, \\
\Omega^{II} &=& \Omega_{\text{rot}}+\Omega_{\text{prec}} &&\text{(line
  II)}. 
\end{array}
\end{equation}
Line I is a consequence of the body's deviation from axisymmetry.
Non-axisymmetry is often discussed in terms of an `equatorial
eccentricity': if the body is approximated as an ellipsoid rotating
about an axis associated with a principal moment of inertia, then it
will look identical after half a rotation period, hence the frequency
$2\Omega_{\text{rot}}$. Line II encodes the precessing motion; recall
that $a/b$ is approximately the wobble angle. The coupling of
precession and rotation causes the frequency of this line to be
shifted upwards by $\Omega_{\text{prec}}$; note that both motions are
in the same sense.  As we shall see later on, this coupling also has
an effect near line I, although it only becomes visible at higher
orders in the precession angle.

\section{Second order contributions to the spectrum}\label{sec:2ndOrder}

Zimmermann's calculation of the first order
contribution to the gravitational radiation from rigid body free
precession \cite{zimmermann80a} involved the expansion parameters $m$
and $\delta$ defined by the somewhat awkward expressions (\ref{m})
and (\ref{delta}), and at the end further truncations were made in terms
of different small quantities. 
Here we will expand in terms of small parameters
that have a more direct physical meaning. 

The wobble angle is only one of several small parameters in the
problem. To set up a second-order expansion, we will first specify a
suitable set of expansion parameters and then estimate the size of
these parameters for realistic neutron stars. These estimates 
will tell us to what order we need to expand in parameters other
than the wobble angle parameter 
$a/b$ to arrive at a sufficiently good approximation for the
waveforms. They will also yield information about the amplitudes of
the different spectral lines at first and second order.

Unlike Zimmermann, who used the general expressions for the quadrupole
radiation (\ref{eq:htt}), our starting point will be 
the closed forms (\ref{eq:closed}) for $h_+$ and $h_\times$. 
There is no significant difference between the two approaches, but 
expansions of the closed formulae will lead to somewhat more elegant 
expressions for the spectral line amplitudes. Overall, our aim is to make the 
derivation and physical meaning of both the first-order and second-order 
results as transparent as possible.
   
\subsection{Expansion parameters}

The parameter $m$ defined by eq.~(\ref{m}) encodes
two pieces of information at once: the deviation from axisymmetry
and the precession angle. To understand the role
played by each, factor $m$ as
\begin{equation}
m = 16\kappa\gamma^2,
\end{equation}
where the new small parameters $\gamma$ and $\kappa$ are defined by
\begin{subequations}
\begin{eqnarray}
\gamma &=& {\frac{I_1}{I_3}}{\frac{a}{b}}, \label{gamma}\\
\kappa &=& {\frac{1}{16}}{\frac{I_3}{I_1}}{\frac{I_2-I_1}{I_3-I_2}}.
\label{kappa} 
\end{eqnarray}
\end{subequations}
The parameter $\gamma$ describes the precession angle while $\kappa$
describes the non-axisymmetry
($I_2-I_1$) relative to the axisymmetric non-sphericity
($I_3-I_2$). Note that in all cases that are of astrophysical interest, 
$I_1/I_3 \simeq 1$; this and the factor of $1/16$ in
$\kappa$ have been included for computational convenience.
The parameter $\delta$ (eq.~\ref{delta}) 
is related to the new parameter $\kappa$; it will be
discussed in subsection C below.

We would like to 
stress that specifying these different expansion parameters from
the outset does not imply a departure from the approach of
\cite{zimmermann80a}. Close inspection of that analysis shows that
$\gamma$ and $\kappa$ are in fact used implicitly to truncate to first
order in $a/b$ at the end of the computation. Introducing them from
the beginning merely facilitates the calculation.

In order to develop appropriate parameters for the second order expansion,
we need to find out what astrophysics is involved in 
the new parameters $\gamma$ and $\kappa$, 
and any other relevant `small' quantities. Note that in realistic cases, 
$\gamma$ and $\kappa$ will not be independent,
since the precession is determined for the most part by the loss of 
axisymmetry. Ideally,
it would be a simple exercise in classical mechanics 
to find the relationship: one could start from a 
non-precessing, axisymmetric body and then study what kind of precessing
motions are induced by various continuous deformations. However, 
we may assume that during a starquake
other factors will be at play, such as torques induced by
partial crustal pinning of vortices in the interior. To estimate their
effect we would have to take recourse to the existing models of
pinning, which seem to be contradicted by the
observations of apparently free precession in pulsars 
\cite{stairs00a,link01a,shabanova01a}. For this reason, we will 
mostly disregard any a priori relation between
$\gamma$ and $\kappa$ and keep the discussion as general
as possible.

\subsection{Astrophysics of precessing neutron
  stars}\label{subsec:astrophysics} 
 
Let us first look into some estimates for oblateness, non-axisymmetry, and
the precession angle.

Oblateness was described by Baym and Pines \cite{baym71a} as
\begin{equation}
{\frac{I_3-I_1}{I_3}} ={\frac{3}{2}}\,\beta\,\epsilon_0.
\label{oblateness}
\end{equation}
Here $\beta$ is the rigidity parameter, which depends on the equation of
state; in \cite{ushomirsky00a} it was derived to be 
\begin{equation}
\beta \simeq 1.6 \times 10^{-5}. \label{rigidity}
\end{equation}
For very young neutron stars, the zero strain oblateness $\epsilon_0$ 
will depend on the shape of the star when the crust first solidified;
subsequent starquakes as well as less violent plastic deformations 
will change it. We will take it to be 
of the order of the ratio of rotational 
kinetic energy to
gravitational binding energy:
\begin{eqnarray}
\epsilon_0 &\simeq& {\frac{\Omega_{\text{rot}}^2 R^3}{G M}} \nonumber\\
           &=& 5.2 \times 10^{-2}\,
               \left({\frac{f_{\text{rot}}}{500\,\,\text{Hz}}}\right)^2,
\label{reference}
\end{eqnarray}
where $f_{\text{rot}}=\Omega_{\text{rot}}/2\pi$. We have set $M=1.4
M_\odot$ and 
$R = 10^6$ cm, which are the values for mass and radius we will use
throughout this paper.

Eqns.~(\ref{oblateness}), (\ref{rigidity}), and (\ref{reference}) lead to 
an oblateness of
\begin{equation}
{\frac{I_3-I_1}{I_3}} \simeq 7.7 \times 10^{-7} 
           \,\left({\frac{f_{\text{rot}}}{500\,\,\text{Hz}}}\right)^2
            \,\left({\frac{\beta}{10^{-5}}}\right). \label{resob}
\end{equation}
The implied range agrees with estimates by Alpar and Pines 
\cite{alpar85a}; see also the discussion in Cutler and Jones \cite{cutler01a}.

To arrive at an estimate of $\kappa$, we need information as to
how a neutron star loses its axisymmetry; here we will 
follow the analysis by Link, Franco, and Epstein 
\cite{link98a}. According to their model,
a starquake is expected to induce fault lines
at angles of 30 to 45 degrees relative to the equator. Crust 
material shifts along the faults and accumulates in mountains with
a typical height of $10^{-2}$ cm. We may assume 
\begin{equation}
I_2-I_1 \sim M_{\text{m}} R^2, \label{assume} 
\end{equation}  
with $M_{\text{m}}$ the mass contained in a 
mountain and $R$ the radius of the neutron star. One has
\begin{equation}
M_{\text{m}} \sim {\frac{\Delta R L}{R^2}} M_{\text{c}},
\label{mountain}
\end{equation}
where $\Delta R$ is the change in equatorial 
radius due to the decrease in oblateness,
$L$ is the fault length, and $M_{\text{c}}$ is the mass of the crust. The 
factor
in front of $M_{\text{c}}$ in (\ref{mountain}) is 
the fraction of the star's area that 
moves with the fault. Assuming an outer crust 
density\footnote{The crust density will increase as one
moves downward. Hence, the use of the outer crust density
to compute the crust mass will lead to an underestimate of $\kappa$,
which in turn will yield an underestimate of the amplitude of line I.} of 
$\rho \sim 10^{10}$ 
$\text{g}/\text{cm}^3$ and 
a crust thickness in the order of $R/10$, we can use the expression 
for $\Delta R/R$ in \cite{link98a} to arrive at
\begin{equation}
{\frac{I_2-I_1}{I_3}} \simeq 2.7 \times 10^{-9}\,{\frac{L}{R}} \,
    \left({\frac{f_{\text{rot}}}{500 \,\,\text{Hz}}}\right)^2 \,
    \left({\frac{t_{\text{q}}}{10 \,\, \text{yr}}}\right) \,
    \left({\frac{t_{\text{age}}}{10^3 \,\, \text{yr}}}\right)^{-1}, 
\label{nonax}
\end{equation}
where $t_{\text{q}}$ is the average time between large starquakes and
$t_{\text{age}}$ 
is the spin-down age. 

We now have 
sufficient information to estimate $\kappa$. 
Using (\ref{kappa}), (\ref{resob}), and (\ref{nonax}), we find
\begin{equation}
\kappa \simeq 2.2 \times 10^{-4} \, {\frac{L}{R}} \,
     \,\left({\frac{\beta}{10^{-5}}}\right)^{-1}
    \left({\frac{t_{\text{q}}}{10 \,\, \text{yr}}}\right) \,
    \left({\frac{t_{\text{age}}}{10^3 \,\, \text{yr}}}\right)^{-1}.
\label{estkappa}
\end{equation}

The most important `small' parameter in our analysis will be the wobble 
angle $\gamma$. For sufficiently small rotation frequencies, the wobble angle
is essentially unconstrained.
For rapidly rotating neutron stars (which are the ones
of interest for gravitational wave detection), Jones
and Andersson \cite{jones01a} propose the following expression for
the maximum wobble angle that can be supported by the crust:
\begin{equation}
\gamma_{\max} \simeq 
1.8 \times 10^{-2}\,\left({\frac{500 \,\,\text{Hz}}{f_{\text{rot}}}}\right)^2
\,\left({\frac{u_{\text{break}}}{10^{-3}}}\right), \label{maxwobble}
\end{equation}
where $u_{\text{break}}$
is the strain at  
which the crust fractures. It is easy to understand
this expression qualitatively; if $f_{\text{rot}}$ is large, so will be the
oblateness and hence the equatorial bulge. The star will then have to
displace more matter while precessing, leading to higher crustal strains.
In \cite{ruderman92a}, the breaking strains of terrestrial materials were
extrapolated to yield estimates in the range $10^{-4} <
u_{\text{break}} < 10^{-2}$, 
although the reliability of such extrapolations is unclear. Note that if the 
breaking strain is near the upper limit of this
range, $u_{\text{break}} \simeq 10^{-2}$, the expression (\ref{maxwobble})
implies an unconstrained precession angle even at a
rotation frequency of 100 Hz.

\subsection{Setting up the second order expansion}

From the estimates
(\ref{resob}), (\ref{nonax}), (\ref{estkappa}) and (\ref{maxwobble})
of the previous subsection, we infer that in the frequency
band of interest,
\begin{equation}
\left|\frac{I_\mu - I_\nu}{I_\rho}\right| \ll \kappa \ll \gamma_{\max}
\end{equation}
for $\mu, \nu, \rho = 1, 2, 3$.

We will choose $\gamma$ as our primary expansion parameter. The approximation
(\ref{waveform}) for the waveforms is of first order in $\gamma$; here we
will go to order $\gamma^2$. However, we will only retain first-order
terms in $\kappa$. The latter will be treated on the same footing
as $\gamma^2$, so that terms in, e.g., $\gamma \kappa$ will be 
discarded.

Considering that quantities of the form $|(I_\mu-I_\nu)/I_\rho|$
are our smallest parameters, we will also take those to be
of order $\gamma^2$.
Accordingly, the quantities $\Gamma_\mu$ defined in
(\ref{deltagamma}) will be approximated as 
$\Gamma_1 \simeq \Delta_2 - \Delta_3$ whenever multiplied 
with $\gamma$ or $\kappa$, and similarly
for $\Gamma_2$, $\Gamma_3$ with cyclic permutation of the indices.

In the expansions of the waveforms, 
linear combinations of $\Delta_\mu$ and $\Gamma_\mu$
defined in (\ref{deltagamma}) will appear, multiplied with powers 
of $\gamma$ and $\kappa$. With the approximations above, 
these can be written as linear 
combinations
of $I_3-(I_1+I_2)/2$ and $I_2-I_1$, the first of which measures oblateness. 
Note that we have 
\begin{equation}
I_2-I_1 \simeq 16\,\kappa\,[I_3-(I_1+I_2)/2] \label{I2-I1},
\end{equation}
the correction being of order $\kappa^2$.
Also, the 
non-axisymmetry parameter $\delta$ defined in (\ref{delta}) can be 
expressed in terms of 
$\kappa$ as
\begin{eqnarray}
\delta &\simeq& 1-\left(1-\frac{I_2-I_1}{I_3-I_2}\right)^{1/2} \nonumber\\
       &\simeq& 1-(1-16 \kappa)^{1/2} \nonumber\\
       &\simeq& 8 \kappa,
\end{eqnarray}
again up to terms in $\kappa^2$. 

We are now ready to calculate the spectrum at second order in the wobble
angle.

\subsection{The spectrum at second order}

The expressions for $R_{i\mu}$ and $\Omega_\mu$ given in 
\cite{zimmermann80a} can be expanded to second order in $\gamma$ and
to first order in $\kappa$. By substituting the results into (\ref{eq:closed})
and truncating appropriately we find the desired second-order expansion of the
waveforms. 

In the expansions, the time dependence of the waveforms is given
entirely in terms of linear combinations of products of trigonometric
functions, with frequencies that are 
integer linear combinations of the
basic frequencies $\Omega_{\text{rot}}$ and
$\Omega_{\text{prec}}$. These trivially 
allow for a Fourier expansion, and after collecting the prefactors
of the simple sines and cosines, we find that 
the second-order waveforms $h^{(2)}_+$ and $h^{(2)}_\times$ are
of the form
\begin{eqnarray}
h^{(2)}_+ &=& \sum_{k=0}^1
    [A^I_{+,k}\, \cos{}(\Omega^I_{2k} t) 
      + A^{II}_{+,k}
           \, \cos{}(\Omega^{II}_{2k} t) ], \nonumber\\ 
h^{(2)}_\times &=& \sum_{k=0}^1
    [A^I_{\times,k}\, \sin{}(\Omega^I_{2k} t) 
      + A^{II}_{\times,k}  \, \sin{}(\Omega^{II}_{2k} t) ],
\label{Fourier2}
\end{eqnarray}
where we defined
\begin{eqnarray}
\Omega^I_{2k} &=& \Omega^I + 2k \, \Omega_{\text{prec}}, \nonumber\\
\Omega^{II}_{2k} &=& \Omega^{II} + 2k \, \Omega_{\text{prec}}. 
\end{eqnarray}
The contributions with frequencies $\Omega^I$ and $\Omega^{II}$ are
the first-order ones in small wobble angle found by Zimmermann 
\cite{zimmermann80a}. 
The second-order contributions appear
as sidelobes to the first order lines.

We find the following non-zero amplitudes. At first order, we 
retrieve Zimmermann's lines in \cite{zimmermann80a}:
\begin{eqnarray}
A^I_{+,0} &=& -{\frac{2}{r}}\,b^2\,(1+\cos{}^2(i))\,(I_2-I_1), \nonumber\\
A^{II}_{+,0} &=& {\frac{1}{r}}\,b^2\,\sin{}(2i)\,
                      [I_3-(I_1+I_2)/2]\,\gamma. 
\end{eqnarray}
To compare these with the amplitudes from (\ref{waveform}), note that
with the approximations we are making, $\Omega_{\text{prec}} \ll
\Omega_{\text{rot}}$, 
so that $b^2 \simeq \Omega^2_{\text{rot}} \simeq
(\Omega_{\text{rot}}+\Omega_{\text{prec}})^2$ 
(indeed, from (\ref{precfreq}) and (\ref{rotfreq}) it can be seen that
the difference between the three is of second order in $\gamma$).
The reason why we find a simpler frequency prefactor stems from the
fact that we have used the closed expressions (\ref{eq:closed}) instead of
the more general formulae (\ref{eq:htt}) as the basis of our computation,
and that the former do not contain any explicit time derivatives.

At second order in the wobble angle, we find a single extra line:
\begin{equation}
A^I_{+,+1} = {\frac{2}{r}}\,b^2\,(1+\cos{}^2(i))\, 
              [I_3-(I_1+I_2)/2]\,\gamma^2. \label{second1}
\end{equation}

The results for the `cross' polarization are quite similar:
\begin{eqnarray}
A^I_{\times,0} &=& -{\frac{4}{r}}\,b^2\,\cos{}(i)\,(I_2-I_1), \nonumber\\
A^{II}_{\times,0} &=& {\frac{2}{r}}\,b^2\,\sin{}(i)\,
                    [I_3-(I_1+I_2)/2]\,\gamma
\end{eqnarray}
at first order in $\gamma$, and
\begin{equation}
A^I_{\times,+1} = {\frac{4}{r}}\,b^2\,\cos{}(i)\,
                      [I_3-(I_1+I_2)/2]\,\gamma^2 \label{second2}
\end{equation}
at second order.

\begin{figure}
\begin{center}
\includegraphics[height=10cm]{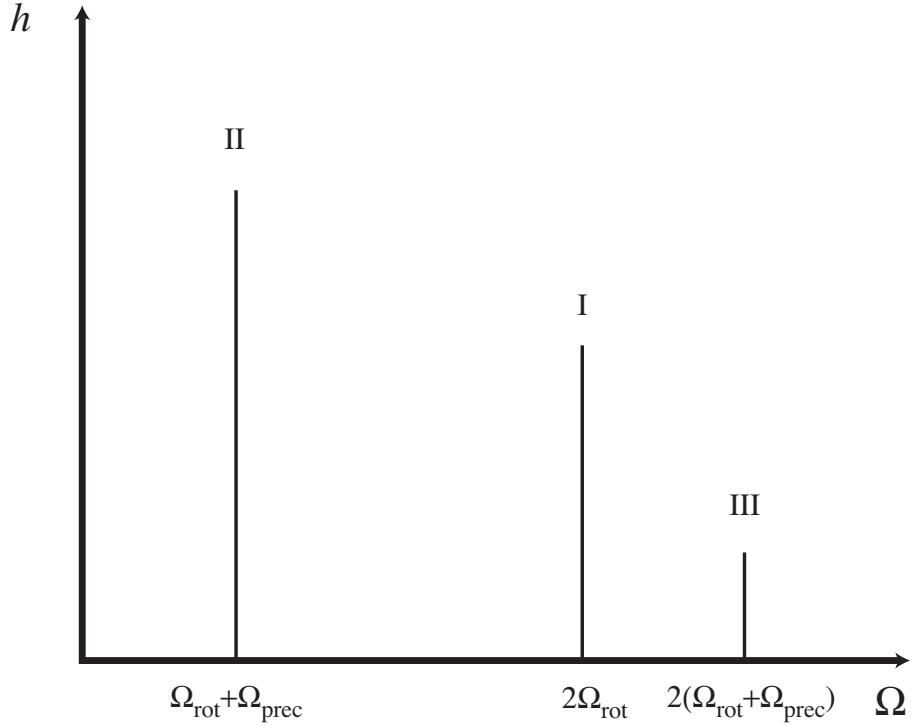}
\end{center}
\caption{A schematic representation of the spectrum at second order 
in the wobble angle. The lines I and II found by Zimmermann et 
al.~\cite{zimmermann79a,zimmermann80a} arise at first order in the
small wobble angle.  
Line I is due to the combination of rotation and non-axisymmetry. Line
II has an amplitude 
proportional to the wobble angle parameter $\gamma$. The amplitude of
line III is
proportional to $\gamma^2$; it appears as a sidelobe to line I.} 
\end{figure}

Henceforth, the second-order line will be referred to as line III. 

Figure 1 gives a schematic representation of the spectrum at second
order in the precession angle.

\section{Detectability and physical content of the
  spectrum}\label{sec:detectability} 

\subsection{Detectability}

In subsection \ref{subsec:astrophysics}, astrophysical estimates 
were obtained for the small parameters governing the problem: the
wobble angle $\gamma$, the deviation from axisymmetry $\kappa$,
and the oblateness. With the expressions for the amplitudes of 
the spectral lines found in the previous subsection, these can now be 
used to determine the detectability of the lines. For concreteness, we 
will be interested
in distances of $r \sim 10$ kpc, rotation frequencies $f_{\text{rot}}
\sim 500$ Hz, and a rigidity parameter $\beta \sim 10^{-5}$. 
Factors of $G$ and $c$ in the amplitudes are reinstated.

The (`cross' polarization) amplitude for line I is of the form
\begin{eqnarray}
A^I &=& -{\frac{G}{c^4}}{\frac{4}{r}}\,b^2\,f(i)\,(I_2-I_1) \nonumber\\
    &=& -I_3{\frac{G}{c^4}}{\frac{64}{r}}\,b^2\,f(i)
\,{\frac{I_3-(I_1+I_2)/2}{I_3}}\,\kappa
\end{eqnarray}
with $f(i)$ some function of order 1 depending on the polarization; 
in the second line we have used
(\ref{I2-I1}). Using the estimates of subsection \ref{subsec:astrophysics} 
and setting
$I_3 = (2/5) (1.4\,M_\odot)(10^6\,\mbox{cm})^2$, we find
\begin{equation}
A^I \simeq 3.3 \times 10^{-28} \,\left({\frac{10\,\,\text{kpc}}{r}}\right)
       \,\left({\frac{f_{\text{rot}}}{500\,\,\text{Hz}}}\right)^4\,
       {\frac{L}{R}}\,
    \left({\frac{t_{\text{q}}}{10 \,\, \text{yr}}}\right) \,
    \left({\frac{t_{age}}{10^3 \,\, \text{yr}}}\right)^{-1}.
\label{nonaxline}
\end{equation}
For line II we have (under the same assumptions):
\begin{eqnarray}
A^{II} &=& I_3 {\frac{G}{c^4}}{\frac{2}{r}}\,b^2\, g(i) \,
                    {\frac{I_3-(I_1+I_2)/2}{I_3}}\,\gamma \nonumber\\
     & \simeq& 4.6 \times 10^{-26} 
              \, \left({\frac{10\,\,\text{kpc}}{r}}\right)\,
                   \left({\frac{f_{\text{rot}}}{500\,\,\text{Hz}}}\right)^4
                   \,\left({\frac{\beta}{10^{-5}}}\right) \,\gamma,
\label{precline}
\end{eqnarray}
where $g(i)$ is again a function of order 1. Finally, for line III:
\begin{equation}
A^{III} \simeq 9.4 \times 10^{-26} 
              \, \left({\frac{10\,\,\text{kpc}}{r}}\right)\,
                   \left({\frac{f_{\text{rot}}}{500\,\,\text{Hz}}}\right)^4
                   \,\left({\frac{\beta}{10^{-5}}}\right) \,\gamma^2.
\label{precline2}
\end{equation}

In Jones and Andersson \cite{jones02a}, detectability of spectral
lines from precessing neutron stars was discussed using estimates of
the sensitivity of interferometric detectors that differ from the ones
we will use. Their paper was written before the more recent analysis
of the expected capabilities of LIGO II by Fritschel 
\cite{fritschel01a,fritschel03a}. 
Here we will assume a LIGO II detector,
optimized for isolated neutron star observations by adding a signal
recycling mirror at the output, the position of which can be changed
macroscopically.  For frequencies between 500 and 1000 Hz, this leads
to a strain sensitivity of $10^{-24}$ $\text{Hz}^{-1/2}$
\cite{fritschel01a,fritschel03a}. 
A one-year integration time would then allow for
the detection of signals with amplitudes as low as
\begin{eqnarray}
h_{\text{min}} &\simeq& 10^{-24}\,\left({\frac{1}{3.15 \times 10^7}}
\right)^{1/2} \nonumber\\
        &\simeq& 2 \times 10^{-28}.
\end{eqnarray}

From (\ref{nonaxline}), we see that line I should be detectable for a spin-down
age much less than $10^3$ years. (Also recall that our estimate for
the magnitude of $\kappa$ should be considered pessimistic.) 
Eq.~(\ref{precline}) indicates that to observe line II, one would need 
$\gamma > 4.3 \times 10^{-3}$. This is
to be compared with the physical upper limit implied by eq.~(\ref{maxwobble}), 
which is $\gamma_{\text{max}} \simeq 1.8 \times 10^{-2}$ for a breaking strain
of $u_{\text{break}} \sim 10^{-3}$.
 
Finally, we come to line III. Eq.~(\ref{precline2}) 
shows that one needs 
\begin{equation}
\gamma > 4.6 \times 10^{-2} \label{needed}
\end{equation}
for the second-order line to be detectable. Now, recall that there is
much uncertainty about the value of the breaking strain of the crust
\cite{ruderman92a}. If, e.g., $u_{\text{break}} \sim 10^{-2}$, then for
the frequency of $f_{\text{rot}} \sim 500$ Hz we are considering, 
the maximum possible wobble angle would be $0.18$
rad. Since we have been assuming a distance of $r \sim 10$ kpc,
this means that line III may well be observable \emph{even for sources
as distant as the galactic center}.

\subsection{Extracting physical information}

We now investigate what information about a neutron star's
shape and motion can be deduced from the spectrum computed in the previous 
section.

If the neutron star is not precessing, then at most one line will be
present in the spectrum, namely the first-order line I associated 
with non-axisymmetry. We will not consider this case any further.

Even if the non-axisymmetry of the neutron star is insignificant, two 
lines are present; both lines II and III are generated purely by the 
precessing motion. The inclination angle $i$ can always be determined
by comparing the two polarizations for two different lines. Introducing 
non-axisymmetry, line I appears,
which is separated from line III by two times the precession frequency.
Let us assume that line III has enough power to be detectable.
We can then distinguish between two cases.

\begin{quote}
(a) The precession period is large compared to the integration time.
In that case we see the precession line II at a frequency $\Omega_{\text{rot}}+
\Omega_{\text{prec}} \simeq \Omega_{\text{rot}}$. 
Even though lines I and III may not be resolvable, the first-order 
non-axisymmetry line I gets modulated by the second-order precession
line III. When studying data, one would know what function to fit the 
time dependence of this modulation by. The unknowns are the precession 
frequency, and the amplitudes of lines I and III. Assuming the fit would
allow one to estimate these quantities with some degree of accuracy, the rest 
of the discussion for 
this case is identical to that of case (b) below.   

(b) If lines I and III can be 
resolved, three distinct lines are seen. The parameters encoding 
the physical information about the neutron star can then be found
as follows. The ratio of the amplitudes of lines III and II is 
proportional to $\gamma$, the proportionality factor being a function
of the inclination angle $i$. Next, one can look at the ratio of the 
amplitudes of lines I and II. This is proportional to $\kappa/\gamma$, where 
the proportionality factor is again a function of $i$; knowing $\gamma$,
we obtain a value for $\kappa$. The frequencies at which the lines occur 
allow us to determine $\Omega_{\text{prec}}$ and
$\Omega_{\text{rot}}$. Now, the expression  
(\ref{precfreq}) for the precession frequency can be written as
\begin{equation}
\Omega_{\text{prec}} \simeq 
\frac{\pi}{2 K(m)}\frac{I_3-(I_1+I_2)/2}{I_3} \Omega_{\text{rot}}.
\end{equation}
Using
$m = 16\kappa\gamma^2$, from $\Omega_{\text{rot}}$ and
$\Omega_{\text{prec}}$ we can 
then compute 
the oblateness parameter
\begin{equation}
{\frac{I_3-(I_1+I_2)/2}{I_3}}.
\end{equation}
\end{quote}

It is important to 
note that, to extract the physical information one is interested 
in (precession angle, deviation from axisymmetry, and oblateness),
\emph{the first-order spectrum would not suffice}. 

We end this section with a brief comment concerning the distance to the
neutron star emitting the radiation. Consider the amplitude of, e.g., 
line I:
\begin{equation}
A^I = -I_3 {\frac{G}{c^4}}{\frac{64}{r}}b^2\,f(i)
\,{\frac{I_3-(I_1+I_2)/2}{I_3}}\,\kappa,
\end{equation}
where $f(i)$ is again some function of the inclination angle of order one.
As we have just seen, knowledge of the amplitudes and frequencies of
the three spectral lines at second order suffices to infer values 
for $i$, $\kappa$, $[I_3-(I_1+I_2)/2]/I_3$, and of course $b^2 \simeq
\Omega_{\text{rot}}^2$. From the expression above, it is then clear that by
inserting an educated guess for the principal moment of inertia $I_3$ (as
we have been doing to assess the detectability of the lines), we 
can obtain a rough estimate of the distance to the source.

\section{Summary and conclusions}\label{sec:conclusions}

We derived the quadrupole gravitational wave spectrum of a rigidly rotating
and freely precessing 
neutron star up to second order in the precession angle. 

Before doing so,
we reviewed Zimmermann's first-order expansion of the waveforms 
\cite{zimmermann80a} . 
We then introduced
new `small' parameters which are more directly related to the physical
quantities of interest, with a view on facilitating the second order 
expansion and its interpretation, and also clarifying the earlier work. These 
parameters correspond to the oblateness, the deviation from axisymmetry, and 
the precession angle of the neutron star.

Next, we made careful astrophysical estimates of physical quantities
characterizing neutron stars. This led to estimates of the `small' parameters,
which justified the use of the wobble angle as the primary 
expansion parameter for a higher-order expansion of the waveforms, putting
the other small parameters on the same footing as the square of the wobble
angle. In addition, the estimates explained why the truncations 
implicitly made by Zimmermann were physically reasonable.

We then used the new small 
parameters to set up an expansion of the waveforms to 
second order in precession angle. The approach taken differed somewhat from 
that of Zimmermann's in that we used as a starting point 
the closed expressions for the specific
case of quadrupole radiation from rigidly rotating and precessing bodies
instead of the general formulae, truncating directly in terms
of parameters with a clear physical interpretation.
Our method does not imply a significant departure from 
Zimmermann's approach, but it does make the calculations 
and the end result more transparent.

At first order, the two spectral lines predicted by Zimmermann were retrieved: 
line I, which is associated with non-axisymmetry, and line II which results 
from the precessing motion. At second order, a single additional line was 
found (line III). 
It appears as a `sidelobe' to line I and is separated from it
by twice the precession frequency. 

On the basis of our estimates for the small parameters, we evaluated the
detectability of the three spectral lines with a LIGO II detector optimized for
isolated neutron star observations \cite{fritschel01a,fritschel03a}, 
assuming a one-year integration time.
For a rotation frequency of 500 Hz and a distance of 10 kpc, together with
some conservative assumptions concerning neutron star structure, we 
found that:
\begin{itemize}
\item Line I should be detectable if the neutron star's spin-down 
age is much less than $10^3$ years;
\item Observability of line II depends on the crustal breaking strain,
which is subject to much uncertainty. However, the line can be observable
even if the breaking strain is near the lower limit of the range proposed
by Ruderman \cite{ruderman92a};
\item If the breaking strain is near the upper end of this range, line III
can also be detectable.
\end{itemize}
Hence, it may well be possible to observe the entire second-order spectrum
of neutron stars as far away as the galactic center. 

Finally, we investigated what gravitational wave observations
might tell us about the neutron star emitting the 
radiation. The two first-order spectral lines found by Zimmermann would not
allow one to infer much about the intrinsic 
physical properties of the source beyond the rotation
and precession frequencies. To separate the precession angle, the deviation
from axisymmetry and the oblateness from given data, at least one 
additional line is needed. Our results provide the missing information:
if the second order line can be seen, then it does become possible
to infer these quantities. 

The direct measurement of these characteristics
would permit an evaluation of the models for neutron star structure
and evolution that led to our estimates of the small parameters
\cite{baym71a,ushomirsky00a,alpar85a,cutler01a,link98a,jones01a,ruderman92a},
in a way that would be impossible by conventional astronomical means.
The detailed implications for neutron star modeling will be 
investigated elsewhere.

Given our assessment of the observability of the spectral lines we found, it 
may be of interest to extend our results by constructing still 
higher-order expansions. 
The scheme presented here can easily be used to compute any number of 
discrete contributions to the quadrupole gravitational wave spectrum of freely 
precessing neutron stars, modeled as rigid bodies.

\section*{Acknowledgements}

It is a pleasure to thank Vijay Chickarmane, Ian Jones, Steinn Sigurdsson, and
Patrick Sutton for useful discussions. I am especially grateful to 
Lee Samuel Finn for valuable guidance as well as for his help
in improving the presentation of this paper. 
This work was supported in part by the Eberly Research Fund of Penn State, 
NSF grants PHY-00-90091 and PHY-00-99559, 
and the Center for
Gravitational Wave Physics, which is funded by the NSF under
cooperative agreement PHY-01-14375.
Finally, the generosity of the Edward M.~Frymoyer Honors Scholarship program
is gratefully acknowledged.

\end{document}